\documentclass[]{article}
\usepackage{emulateapj}
\usepackage{psfig}

\begin{document}

\title{Photon Propagation Around Compact Objects and the Inferred
Properties of Thermally Emitting Neutron Stars}

 \author{Dimitrios Psaltis, Feryal \"Ozel, and Simon DeDeo}
 \affil{Harvard-Smithsonian Center for Astrophysics, 60 Garden St.,
 Cambridge, MA 02138;\\ dpsaltis, fozel, sdedeo@cfa.harvard.edu}

\slugcomment{To appear in {\em The Astrophysical Journal\/}}

\begin{abstract}
  Anomalous X-ray pulsars, compact non-pulsing X-ray sources in
  supernova remnants, and X-ray bursters are three distinct types of
  sources for which there are viable models that attribute their X-ray
  emission to thermal emission from the surface of a neutron star.
  Inferring the surface area of the emitting regions in such systems
  is crucial in assessing the viability of different models and in
  providing bounds on the radii of neutron stars. We show that the
  spectroscopically inferred areas of the emitting regions may be
  over- or under-estimated by a factor of $\lesssim 2$, because of the
  three-dimensional geometry of the system and general relativistic
  light deflection, combined with the effects of phase averaging. Such
  effects make the determination of neutron-star radii uncertain,
  especially when compared to the $\sim 5$\% level required for
  constraining the equation of state of neutron-star matter. We also
  note that, for a given spectral shape, the inferred source
  luminosities and pulse fractions are anticorrelated because they
  depend on the same properties of the emitting regions, namely their
  sizes and orientations. As a result, brighter sources have on
  average weaker pulsation amplitudes than fainter sources. We argue
  that this property can be used as a diagnostic tool in
  distinguishing between different spectral models. As an example, we
  show that the high inferred pulse fraction and brightness of the
  pulsar RXS~J1708$-$40 are inconsistent with isotropic thermal
  emission from a neutron-star surface. Finally, we discuss the
  implication of our results for surveys in the soft X-rays for young,
  cooling neutron stars in supernova remnants and show that the
  absence of detectable pulsations from the compact source at the
  center of Cas~A (at a level of $\gtrsim 30$\%) is not a strong
  argument against its identification with a spinning neutron star.
\end{abstract}


\section{INTRODUCTION}

Neutron stars in nature appear in various different flavors. They were
first unambiguously discovered in radio wavelengths as
rotation-powered pulsars and later on, in X-rays, as accretion-powered
pulsars and bursters. Such systems, however, constitute only a small
fraction of the expected number of neutron stars in the galaxy, as
inferred from estimates of supernova rates (e.g., Kaspi 2000). As a
result, ongoing searches exist for detecting neutron stars in
different manifestations, e.g., as isolated stars accreting from the
interstellar medium (e.g., Belloni et al.\ 1997) or simply cooling by
thermal emission (e.g., Pavlov et al.\ 1996). Different emission
mechanisms are thought to operate in different types of neutron stars
and possibly even between different wavelength bands in a given
system. For at least three distinct classes of neutron stars, which we
discuss below, viable models exist in which radiation emerging
directly from their surfaces is responsible for their high-energy
spectra.

A number of slow X-ray pulsars, often called the anomalous X-ray
pulsars (AXPs; Mereghetti \& Stella 1995) form a distinct class of
neutron stars with soft X-ray spectra but no radio or optical
counterparts. Because of their high spin-down rates and association
with supernova remnants (SNRs) they are thought to be young and
related to the soft $\gamma$-ray repeaters, in their quiescent states
(see Hurley 2000). It is still an open question whether such objects
are neutron stars accreting from a fossil disk (e.g., van Paradijs et
al.\ 1996; Chaterjee, Narayan, \& Hernquist 2000), are powered by
magnetospheric emission (e.g., Thompson \& Duncan 1995, 1996), or even
emitting thermally (Heyl \& Hernquist 1998).

Another class of objects, potentially related to the AXPs are the
compact, non-pulsing, soft X-ray sources that are being discovered
within SNRs. The most recent of such compact objects is the central
source in Cas A discovered with the {\em Chandra X-ray Observatory\/}
(Tananbaum 1999; see also Pavlov et al.\ 2000; Chakrabarty et al.\ 
2000).  Their spectra and luminosities are typical of what is expected
for cooling, young neutron stars (see also Pavlov et al.\ 1996; Zavlin
et al.\ 1998, 1999). However, the absence of coherent pulsations in
their X-ray brightness allows for the possibility that they are black
holes accreting from fallback material (see, e.g., Pavlov et al.\ 
2000; Chakrabarty et al.\ 2000).

Finally, weakly-magnetic ($\lesssim 10^{10}$~G) accreting neutron
stars often show thermal emission from their surfaces during
thermonuclear flashes, the so-called Type I X-ray bursts (see Lewin,
van Paradijs, \& Taam 1996). In six bursters, relatively coherent
oscillations are detected during the bursts, at frequencies $\sim
300$~Hz (see, e.g., Strohmayer et al.\ 1996). The coherence and
stability of the oscillations suggests that they occur at the
neutron-star spin frequencies and are caused by the non-uniform
pattern of burning in their surface layers.

Understanding the properties of the X-ray emission from such thermally
emitting neutron stars, especially in connection to the presence or
absence of detectable coherent pulsations, is crucial in assessing the
viability of different models. Moreover, comparing model spectra to
observations offers the possibility of measuring the radii of neutron
stars and hence constraining the properties of neutron-star matter
(see Lattimer \& Prakash 2000 for a recent discussion). Recently,
there has been significant progress in calculating spectra emerging
from neutron-star atmospheres with various compositions (see, e.g.,
Pavlov et al. 1996; Rajagopal, Romani, \& Miller 1997). Constraints on
the amplitudes of oscillations from compact stars with non-uniform
surface emission have also been studied in connection to the observed
oscillation amplitudes in cooling radio pulsars (Page 1995), bursters
(see, e.g., Miller, \& Lamb 1998; Weinberg, Miller, \& Lamb 2000), and
AXPs (DeDeo, Psaltis, \& Narayan 2000).

In this article, we address a number of issues related to the emission
from a spinning neutron star with a non-uniform surface brightness.
In particular, in \S2 we study the effects of the three-dimensional
geometry of the systems, phase averaging, and the general relativistic
deflection of light. In \S3, we demonstrate that, when the radiation
emerges from a localized surface area on a rotating star, the
radiation flux reaching an observer at infinity may be substantially
different compared to the case of isotropic, spherically symmetric
emission from a compact star with the same surface area (see also
Zavlin, Shibanov, \& Pavlov 1995; Leahy \& Li 1995). As a result, when
such effects are not taken explicitly into account, the hot-spot
sizes, inferred from the observed fluxes and temperatures, may be
significantly over- or under-estimated. We show that
general-relativistic deflection typically reduces this discrepancy,
depending on the compactness of the neutron star.

We also note that the surface area on the neutron star, from which the
localized emission emerges, determines both the flux that reaches the
observer at infinity and the amplitude of pulsations at the stellar
spin frequency. In fact, for stars with the same local surface
temperature and emerging spectrum, an anticorrelation is expected
between the luminosity and pulsation amplitude.  Therefore,
simultaneous consideration of these two properties for a given system
can offer strong constraints on the operating emission mechanisms. In
\S4, we investigate this property and its implications for the
observations of compact X-ray sources with and without detectable
pulsations.

\section{THE BRIGHTNESS OF A SPINNING NEUTRON STAR}

In this section, we calculate the phase-averaged radiation flux that
reaches an observer at infinity from the surface of a spinning neutron
star (or other compact object) with a non-uniform surface brightness,
following the procedure outlined by Pechenick, Ftaclas, \& Cohen
(1983; see DeDeo et al.\ 2000 for the details of our implementation).
Throughout this article, we neglect any interaction of photons with
matter between the surface of the neutron star and the observer, as
well as any polarization effects (see, e.g., Shaviv, Heyl, \& Lithwick
1999). We also assume that the star is slowly rotating, so that its
spacetime is described by the Schwarzschild metric. Finally, we set
$c=G=1$, where $c$ is the speed of light and $G$ is the gravitational
constant.

We specify, as a boundary condition, the specific intensity
$I(\theta,\phi,\theta')$ of radiation emerging from the neutron-star
surface, evaluated at the local rest frame and integrated over all
photon energies. The polar coordinates $(\theta,\phi)$ determine the
position on the stellar surface with respect to the rotation axis,
while the angle $\theta'$ is measured locally with respect to the
radial direction. Hereafter, we assume that the emission is isotropic
and hence that the specific intensity is independent of $\theta'$.
This choice leads to the strongest effects of general relativistic
light deflection, even though it is not necessarily the most
appropriate for thermal emission from a neutron star (see, e.g.,
Zavlin et al.\ 1998). Calculating in detail the beaming of radiation
requires the knowledge of the temperature stratification of the
neutron-star atmosphere and the solution of the resulting
radiative-transfer problem, which is beyond the scope of this paper
but will reported elsewhere.

The brightness distribution on the neutron star surface depends on the
emission process under consideration (i.e., thermal cooling versus
localized nuclear burning), the surface profile of its magnetic field
(see, e.g., Heyl \& Hernquist 1998), and the possible presence of
lateral metallicity gradients (see, e.g., Pavlov et al.\ 2000). For
the purposes of our analysis, considering a geometry of two antipodal
hot spots of variable size with uniform surface brightness captures
the relevant properties of these different processes. Following the
notation used for radio pulsars (e.g., Lyne \& Graham-Smith 1990), we
denote the half-opening angle of each spot by $\rho$, its angular
distance from the rotation pole by $\alpha$, and the emerging constant
specific intensity by $I_{\rm NS}$.

The flux measured by an observer at distance $d$, whose polar
coordinates with respect to the stellar rotation axis are
denoted by $(\beta,\Phi)$, is given by (Pechenick et al.\ 1983, eq.~[3.15])
\begin{eqnarray}
  F_{\infty}(\beta,\Phi)&=& I_{\rm NS}\left(\frac{R_{\rm NS}}{d}\right)^2
        \left(\frac{M_{\rm NS}}{R_{\rm NS}}\right)^2\nonumber\\ 
        & &\times\left(\sqrt{-g_{00}}\right)^4 \int_{0}^{x_{\rm
        max}} h[\theta(x,\beta);\rho,\theta_0]x dx\;.
\label{eq:flux_phi}
\end{eqnarray}
Here, $M_{\rm NS}$ and $R_{\rm NS}$ are the neutron-star mass and
radius, $g_{00}\equiv -(1-2M_{\rm NS}/R_{\rm NS})$ for a Schwarzschild
spacetime, $x_{\rm max}\equiv(R_{\rm NS}/M_{\rm NS})/ \sqrt{-g_{00}}$,
and the function $h(\theta; \rho, \theta_0)$ is defined in Pechenick
et al.\ (1983, eq.~[3.3]). The angle $\theta_0$ measures the distance
on the stellar surface between the center of one of the hot spots and
the direction to the observer and is given by
\begin{equation}
 \cos\theta_0=\sin\alpha\sin\beta\cos\Phi+\cos\alpha\cos\beta\;.
\label{eq:th0}
\end{equation}
Equation~(\ref{eq:flux_phi}) is simply the integral of specific
intensities, at the distance of the observer, over the impact
parameters $b=xM_{\rm NS}$ of rays that are parallel at radial
infinity. The angle $\theta(x,\beta)$ at which each parallel ray
intersects the stellar surface is found by integrating the photon
trajectory from radial infinity to the stellar surface, i.e.,
\begin{equation}
  \theta=\int_{0}^{M_{\rm NS}/R_{\rm NS}} [x^{-2}-(1-2u)u^2]^{-1/2}du
\label{eq:th}
\end{equation}
(Pechenick et al.\ 1983). 

The average flux, measured by an observer at infinity, averaged over a
time longer than the rotational period of the neutron star, simply is
\begin{equation}
F_\infty = \pi I_{\rm NS} \left(\sqrt{-g_{00}}\right)^2
        A(\alpha,\beta,\rho,p)
        \left(\frac{R_{\rm NS}}{d}\right)^2\;,
\label{eq:Finfty}
\end{equation}
where $p\equiv R_{\rm NS}/2M_{\rm NS}$ and we have defined
\begin{equation}
A(\alpha,\beta,\rho,p)\equiv
        \left(\frac{-g_{00}}{8\pi^2 p^2}\right) 
        \int_{0}^{2\pi} d\Phi
        \int_{0}^{x_{\rm max}} h[\theta(x,\beta);\alpha,\theta_0]x dx\;.
\label{eq:Aint}
\end{equation}
When the local radiation spectrum emerging from the stellar surface is
that of a blackbody of temperature $T_{\rm NS}$, then $\pi I_{\rm NS}=
\sigma T_{\rm NS}^4$, where $\sigma$ is the Stefan-Boltzmann constant.

If the emission were spherically symmetric, an observer at radial
infinity would measure a specific intensity $I_\infty= (-g_{00})
I_{\rm NS}$ and a blackbody temperature $T_\infty = (\sqrt{-g_{00}})
T_{\rm NS}$. In the absence of any prior information about the opening
angle of the hot spots and their orientation, an observer at infinity
would therefore infer for the emitting region a surface area
\begin{equation}
  S_\infty \equiv \frac{4 d^2 F_\infty}{I_\infty}
        =4\pi d^2 \frac{F_\infty}{\sigma T_\infty^4}\;.
  \label{eq:Soo}
\end{equation}
It is then customary to assume a given neutron-star mass and radius
and correct for the effect of gravitational redshifts as (see, e.g.,
Lattimer \& Prakash 2000)
\begin{equation}
S_{\rm inf}\equiv (-g_{00})S_\infty=
        (-g_{00})4\pi d^2 \frac{F_\infty}{\sigma T_\infty^4}\;.
\label{eq:Sinf}
\end{equation}
Given that the real surface area of two polar caps of half-opening
angle $\rho$ is $S_{\rm pc}=4\pi(1-\cos\rho)R_{\rm NS}^2$, we
conclude that the error in the estimate of the emitting area is
\begin{equation}
\frac{S_{\rm inf}}{S_{\rm pc}}=
        \frac{1}{1-\cos\rho}A(\alpha,\beta,\rho,p)\;.
\label{eq:eS}
\end{equation}
As a consequence, estimating the radius of each of the two polar caps
according to $2\pi D_{\rm inf}^2\equiv S_{\rm inf}$, leads to an error
with respect to the real polar cap radius $D_{\rm pc}=\rho R_{\rm NS}$
equal to
\begin{equation}
\frac{D_{\rm inf}}{D_{\rm pc}}
        =\sqrt{\frac{2A(\alpha,\beta,\rho,p)}{\rho^2}}\;.
\end{equation}
Finally, assuming that the whole neutron star is emitting uniformly
and using equation~(\ref{eq:Sinf}) to infer its radius, results in an
error equal to
\begin{equation}
\frac{R_{\rm inf}}{R_{\rm NS}}=A(\alpha,\beta,\rho,p)\;.
\end{equation}

\section{RESULTS}

\subsection{Uncertainties in the Inferred Surface Areas of the Emission Regions}

We present in this section the uncertainties introduced to the
estimates of the spectroscopically inferred polar-cap surface areas by
the three-dimensional geometry of the problem, the effects of phase
averaging, and the general relativistic light deflection.

Figures~1 and 2 show the ratio of the inferred to the intrinsic
surface areas of the polar caps, for different opening 

\vbox{ \centerline{ \psfig{file=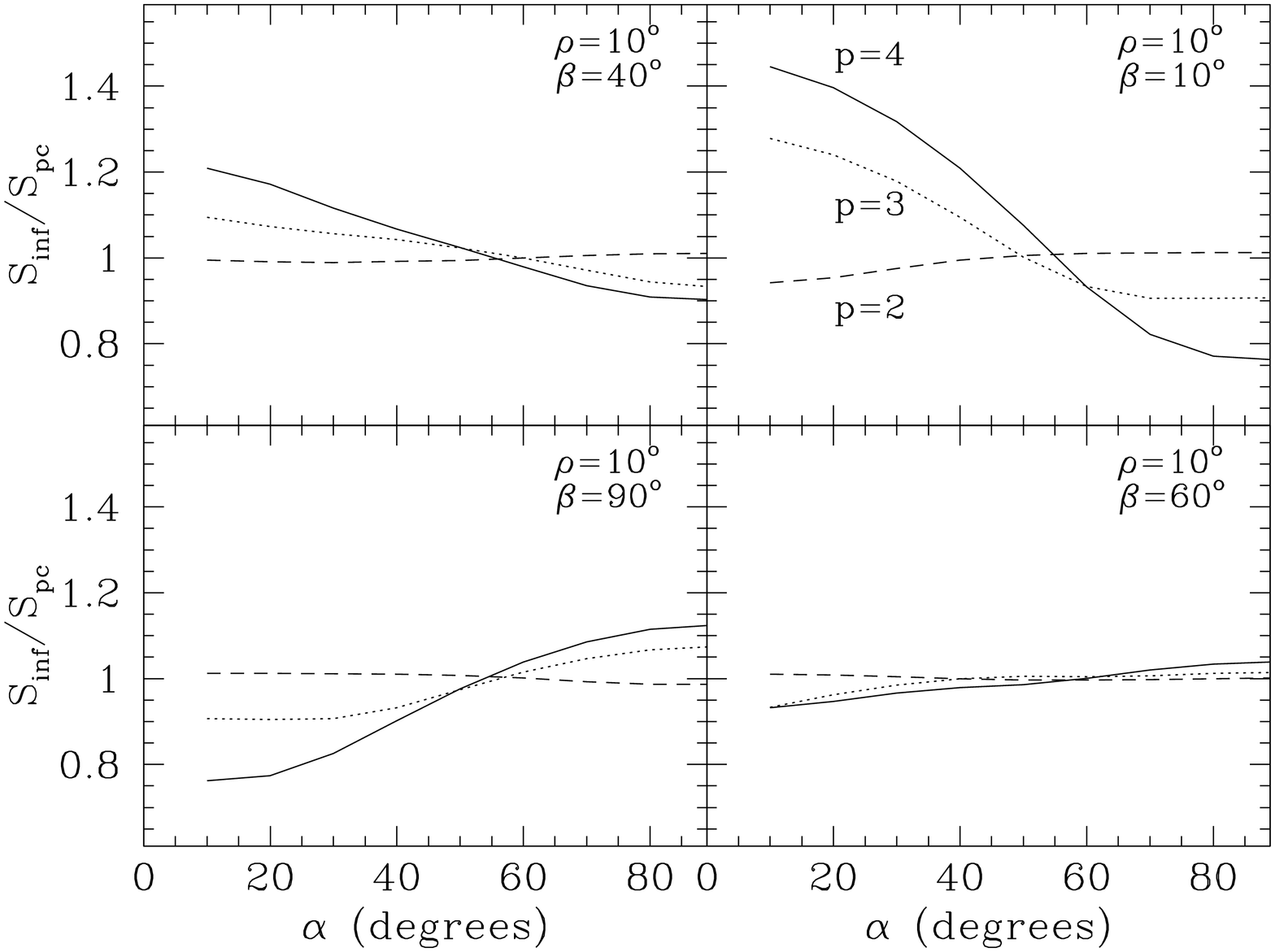,angle=-90,width=9.0truecm} }
    \figcaption[]{\footnotesize Ratio of spectroscopically inferred
    polar-cap surface area $S_{\rm inf}$ to their intrinsic area
    $S_{\rm pc}$, as a function of the orientation of the polar caps
    ($\alpha$), for various values of the observer inclination
    ($\beta$) with respect to the rotational axis and for different
    neutron-star radii. The opening angle $\rho$ of each polar cap
    is held fixed at 10 degrees.}}
\vspace*{0.5cm}
\vbox{ \centerline{ \psfig{file=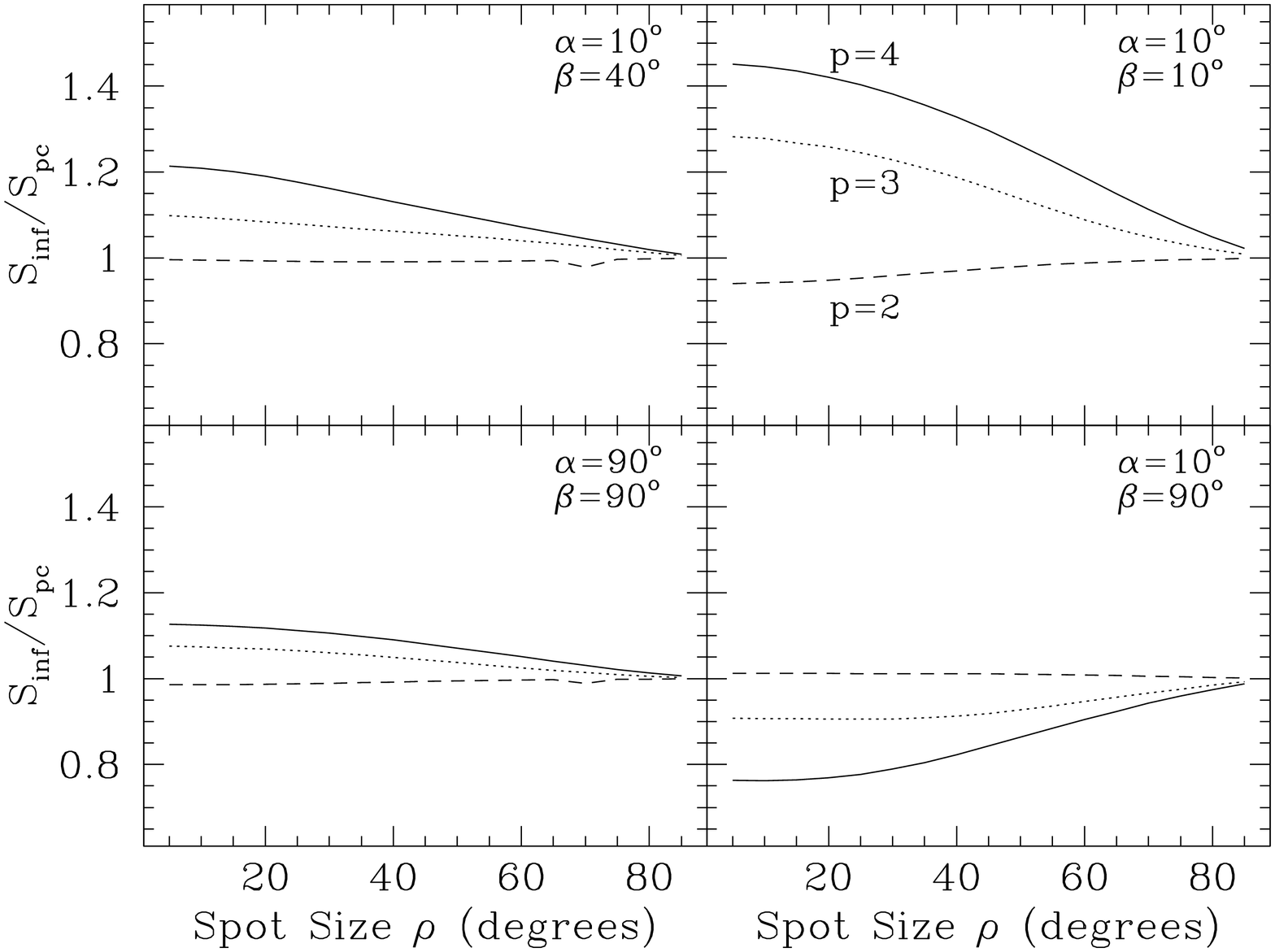,angle=-90,width=9.0truecm} }
    \figcaption[]{\footnotesize Same as in Figure~1 but as a function 
    of the opening angle $\rho$ of each polar cap.}}
\vspace*{0.5cm}

\noindent angles,
neutron-star radii, and orientations of the rotation axis with respect
to the polar caps and the observer. For neutron-stars that are not
very relativistic ($p=4$), the inferred surface areas can be
significantly over- or under-estimated, depending on the relative
orientations of the polar caps and the observer. In order to
understand this effect, we calculate explicitly the ratio $S_{\rm
inf}/S_{\rm pc}$ for the limiting case of a Newtonian star, an
infinitesimally small emitting area, and two specific orientations.

When $\alpha=\beta=0^\circ$, one polar cap always appears at the
geometric center of the stellar disk, and the flux measured at
infinity is time independent. In this case, $\cos\theta_0=1$ and, by
symmetry,
\begin{equation}
h[\theta(\chi,\beta);\rho\rightarrow 0,\theta_0=0^\circ]=2\pi\;.
\end{equation}
Using these values in evaluating the integral~(\ref{eq:Aint}), we
obtain
\begin{equation}
A(\alpha=\beta=0^\circ, \rho\rightarrow 0, p\rightarrow \infty)
   =2(1-\cos\rho)
\end{equation}

\vbox{ \centerline{ \psfig{file=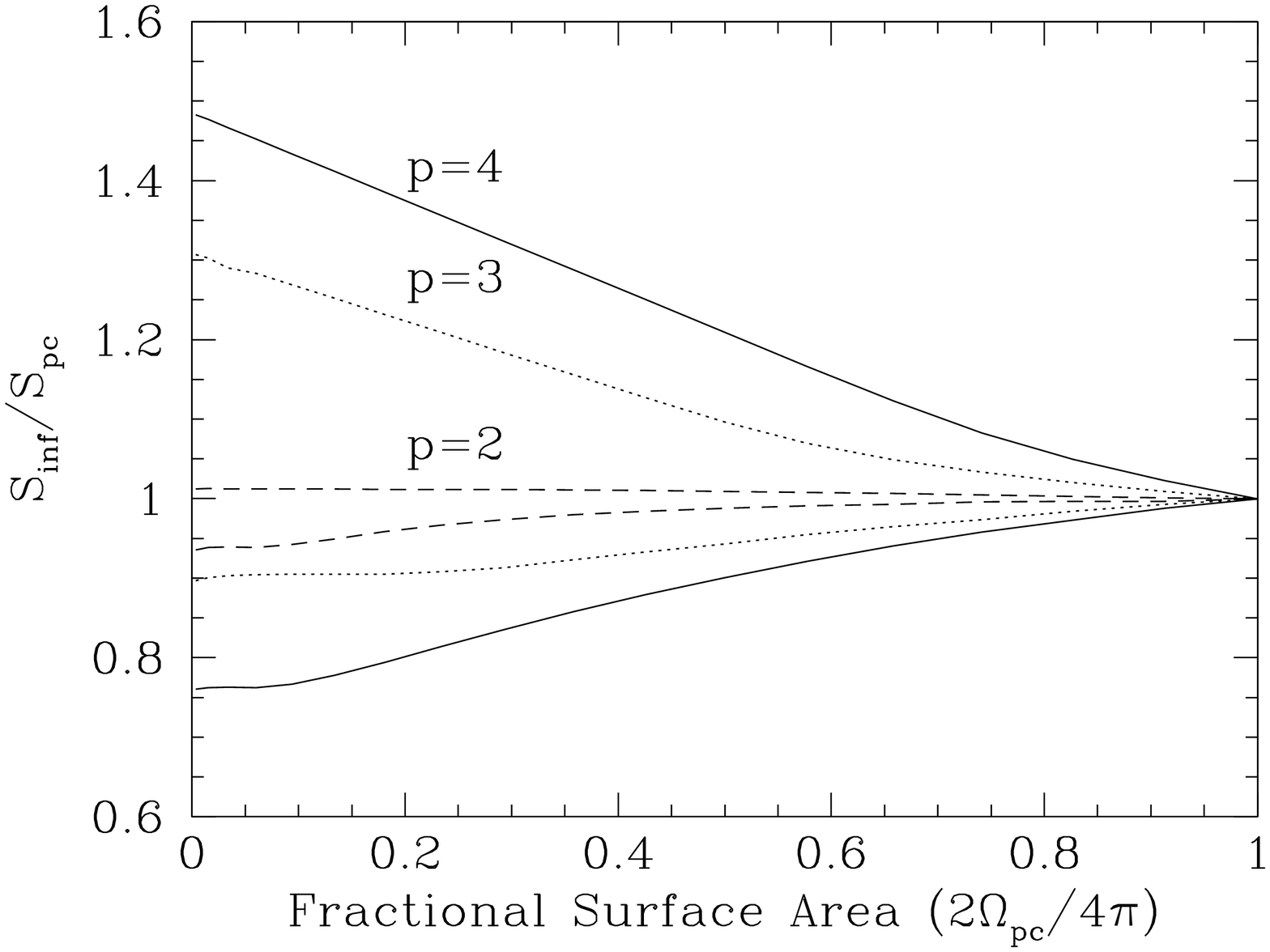,angle=-90,width=9.0truecm} }
    \figcaption[]{\footnotesize Bounds on the ratio of the inferred
    $S_{\rm inf}$ to the intrinsic size $S_{\rm pc}$ of emitting polar
    caps, for different emitting fractions of the stellar surface and
    neutron-star radii.}}
\vspace*{0.5cm}

\noindent 
which gives $S_{\rm inf}/S_{\rm pc}=2$. When, on the other hand,
$\alpha=0^\circ$ and $\beta=90^\circ$, $\cos\theta_0=0$, the polar
caps are being viewed only at grazing angles. In this case,
\begin{equation}
 h[\theta(\chi,\beta);\rho\rightarrow 0, \theta_0=90^\circ]\chi\rightarrow 0,
\end{equation}
and
\begin{equation} 
A(\alpha=0^\circ, \beta=90^\circ,\rho\rightarrow 0, p\rightarrow \infty)
   \rightarrow 0
\end{equation}
and hence $S_{\rm inf}/S_{\rm pc}\rightarrow 0$. As a result, for a
Newtonian star and an infinitesimally small emitting area,
\begin{equation}
0\le \frac{S_{\rm inf}}{S_{\rm pc}} \le 2\;.
\end{equation}

For increasingly more compact neutron stars, i.e., for decreasing
values of $p$, the error in the estimate of the emitting area
decreases substantially, reaching $\lesssim 5\%$ when $p=2$. This is
the result of the strong gravitational light bending near the neutron
star surface, which efficiently redistributes the emitting photons to
almost all directions of propagation, mimicking the spherically
symmetric case.

Figure~3 shows the maximum and minimum of the ratio $S_{\rm
  inf}/S_{\rm pc}$ for all possible orientations of the hot spot and
the observer, as a function of the fractional emitting surface area
($S_{\rm pc}/4\pi R_{\rm NS}^2=2\Omega_{\rm pc}/4\pi$) of the
neutron-star surface, for different neutron-star radii. As expected,
in all cases, the maximum corresponds to $\alpha=\beta=0^\circ$ and
the minimum to $\alpha=0^\circ$ and $\beta=90^\circ$, and they both
converge to unity as $\Omega_{\rm pc}\rightarrow 2\pi$. However, for
small polar caps and typical neutron-star masses and radii, there is a
factor of $\lesssim 2$ spread in the systematic uncertainty in the
estimated area of the emitting region.

\subsection{Trends Between the Pulse Fractions and Luminosities of Thermally
Emitting Neutron Stars}

The X-ray brightness of a spinning neutron-star with a non-uniform
surface emission shows pulsations at the stellar spin frequency and
its harmonics. Just as in the case of the flux observed at infinity,
the amplitude of pulsations as well as the harmonic structure also
depend strongly on the brightness distribution on the stellar surface
and the degree of gravitational light bending. Therefore, inevitable

\vbox{ \centerline{ 
\psfig{file=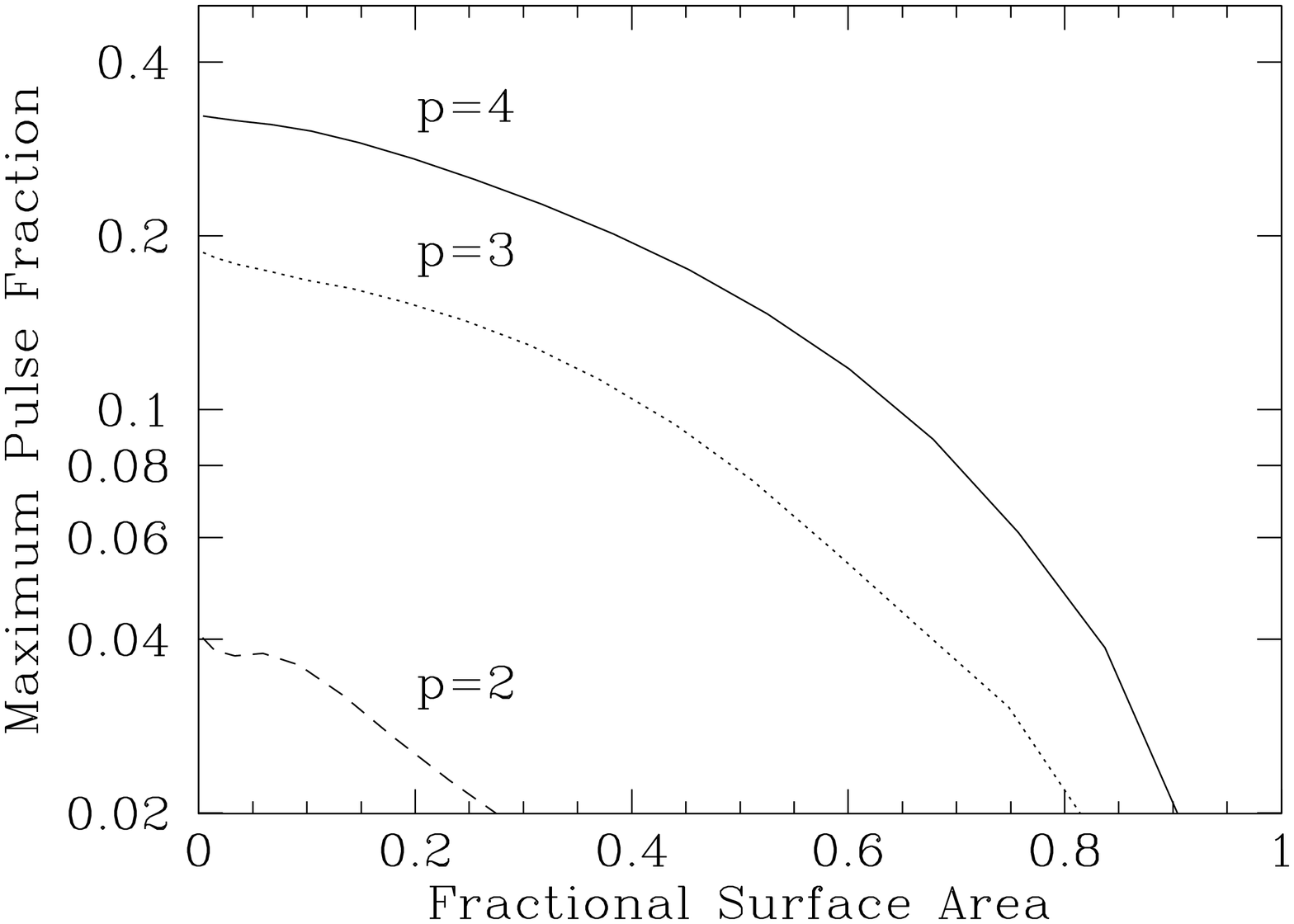,angle=-90,width=9.0truecm} }
  \figcaption[]{\footnotesize Maximum pulse fraction of the brightness
    of a spinning neutron star, for different emitting fractions of
    the stellar surface and neutron-star radii.}}  
\vspace*{0.5cm}
\vbox{ \centerline{
\psfig{file=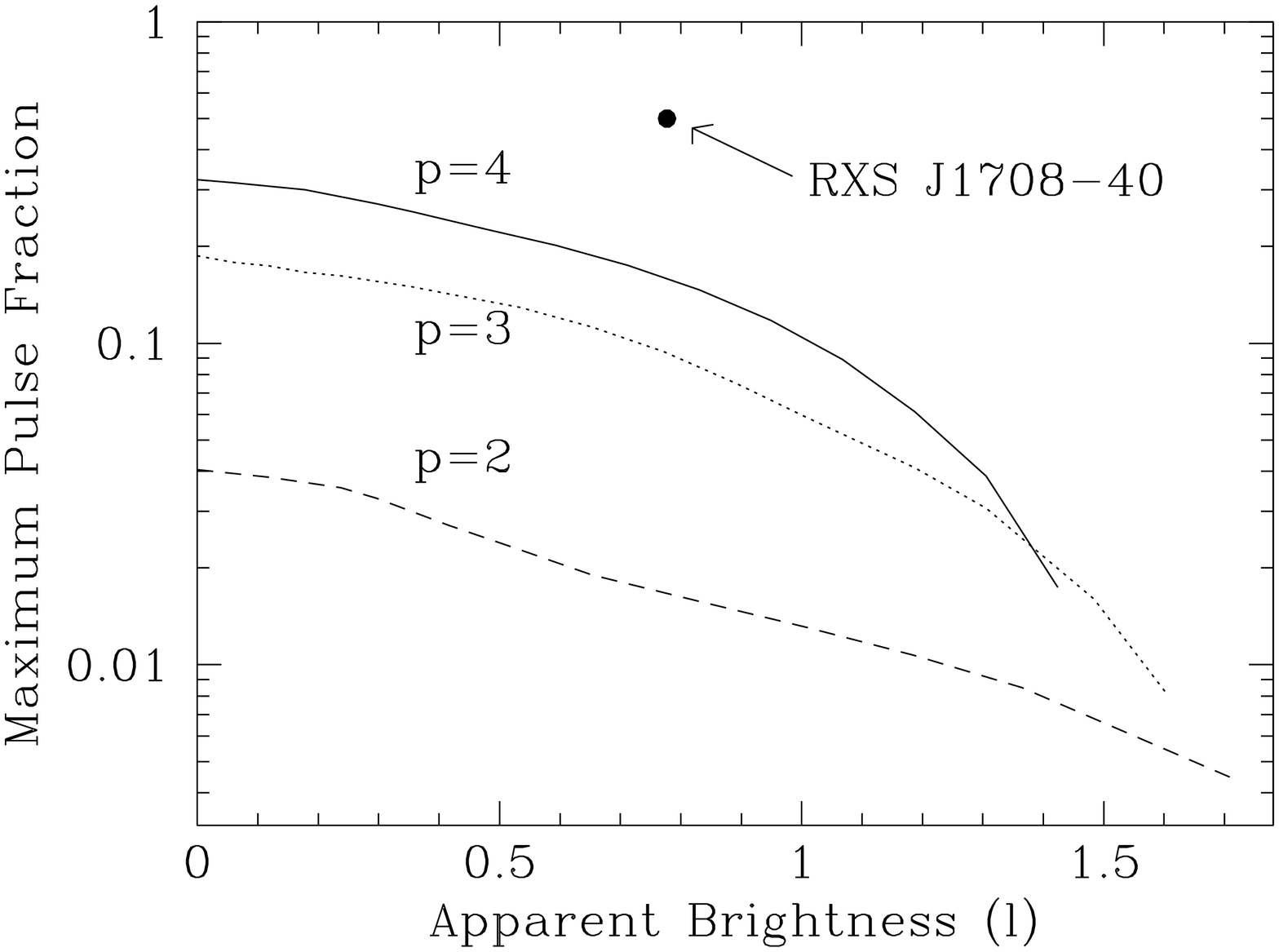,angle=-90,width=9.0truecm} }
  \figcaption[]{\footnotesize Maximum allowed pulse fraction from a
    spinning neutron star as a function of its apparent brightness
    [$l=(4\pi d^2 F_\infty/10^{36}$~erg~s$^{-1})/(T_\infty/$0.5~keV)]
    inferred by an observer at infinity, assuming that the emission is
    spherically symmetric and without applying any redshift
    corrections. The local radiation spectrum is assumed to be that of
    a blackbody and the radius of the neutron star is fixed at 10~km.
    Different relativity parameters correspond to different
    neutron-star masses. The data points correspond to the X-ray pulsar
    RXS~J1708$-$40.}}  \vspace*{0.5cm}

\noindent trends exist between the pulse fractions and the brightness of a
source, as we discuss below.

For the case of two antipodal polar caps discussed here, the amplitude
of pulsations from a neutron star of a given radius decreases with
increasing polar-cap surface area and increasing mass. This is shown
in Figure~4, where the maximum pulse fraction, defined as
\begin{equation}
PF\equiv\frac{F(\Phi)\vert_{\rm max}-F(\Phi)\vert_{\rm min}}
   {F(\Phi)\vert_{\rm max}+F(\Phi)\vert_{\rm min}}\;,
\end{equation}
which occurs for an orthogonal rotator ($\alpha=\beta=90^\circ$), is
plotted against the fractional surface area of the emitting region,
for different neutron-star radii.

\begin{figure*}[t]
\centerline{ \psfig{file=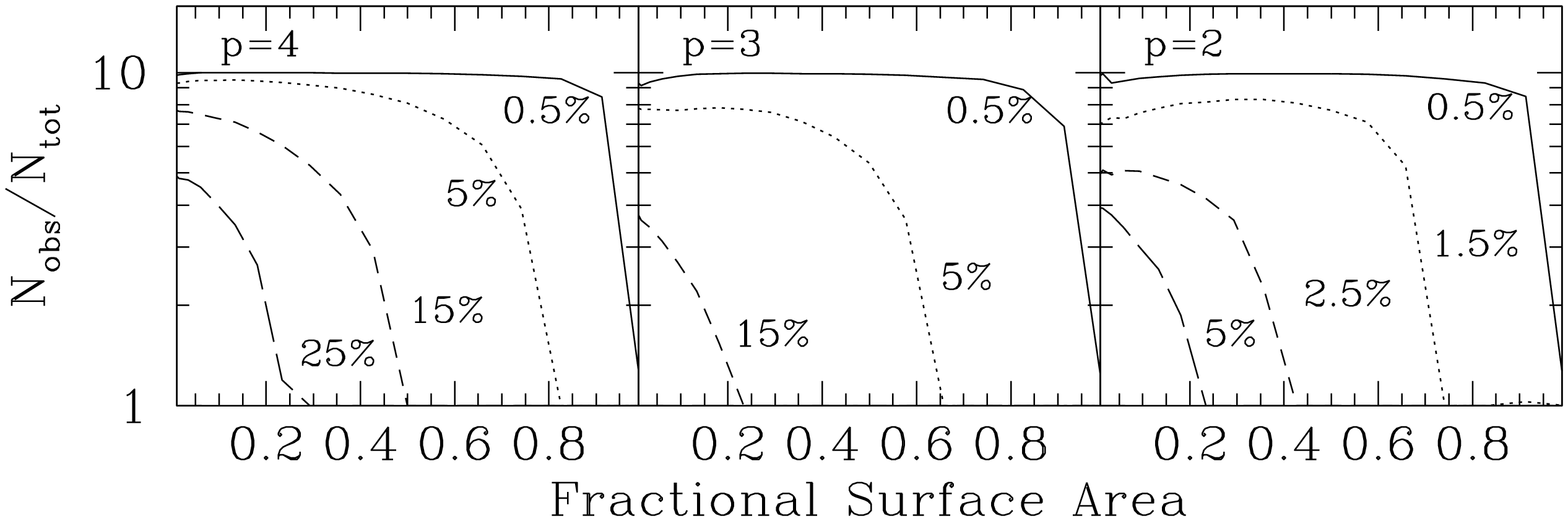,angle=-90,width=14.0truecm,height=6.0truecm} }
    \figcaption[]{\footnotesize Fraction $N_{\rm obs}/N_{\rm tot}$ of
    observed systems with pulse fraction larger than a given threshold
    as a function of the fractional size of the polar caps, for
    different neutron-star radii.}
\end{figure*}

As it has already been discussed extensively in the literature, for
the isotropic beaming we use here, general relativistic light bending
suppresses the pulse fraction below $\sim 35\%$, even for
infinitesimally small polar-cap sizes. As a result, detection of a
pulsation from a neutron star with a larger amplitude severely
constrains any such models of emission from the stellar surface.
However, these constraints may become even tighter if the observed
brightness of the neutron star is also taken into account. For the
same local radiation spectrum, small polar-cap sizes correspond to
large pulse fractions but weak radiation fluxes and vice versa. As a
result, in this case, the maximum pulse fraction for a bright source
is significantly smaller compared to the maximum pulse fraction for a
faint source.

Inferring the fractional surface area of the emitting region, and thus
the source brightness, requires an a priori knowledge of the
orientation of the polar caps and the observer with respect to the
rotation axis, which are almost always unknown.  For example, a given
observed source brightness may be the result of a small polar cap
viewed from a favorable orientation ($\alpha\sim\beta\sim 0^\circ$) or
of a larger polar cap viewed from an unfavorable orientation
($\alpha\sim 0^\circ$, $\beta\sim 90^\circ$). Therefore, since the
observed brightness does not have an one-to-one correspondence with
the fractional emitting surface area, the constraints plotted in
Figure~4 cannot be compared directly with observations. However,
although both configurations in the above example may produce the same
pulse-average flux as measured by the observer, the first
configuration typically produces a significantly smaller pulse
fraction. As a result, the combination of these two properties, namely
the source brightness and pulse fraction, provide us with a useful
diagnostic tool, as we discuss below.

We can simultaneously account for all the above effects if we assume a
particular model of the local radiation spectrum and search for the
maximum pulse fraction as a function of the brightness of the source
measured by an observer at infinity. As an example, we assume that the
local radiation spectrum at each point on the polar caps is that of a
blackbody of temperature $T_{\rm NS}$ with isotropic beaming.  An
observer at infinity, would observe a radiation flux $F_\infty$, and
making the assumption that the emission is spherically symmetric,
would infer a luminosity
\begin{equation}
L_\infty = 4\pi d^2 F_\infty\;.
\end{equation}
According to equation~[\ref{eq:Finfty}], this luminosity is
\begin{equation}
   L_\infty = 4\pi R_{\rm NS}^2\sigma T_{\rm NS}^4 
   \left(\sqrt{-g_{00}}\right)^2 A(\alpha,\beta,\gamma,r)\;,
\end{equation}
where $T_{\rm NS}=T_\infty/(\sqrt{-g_{00}})$. We define 
an apparent brightness as
\begin{equation}
l\equiv \left(\frac{4\pi d^2 F_\infty}{10^{36}~\mbox{erg s}^{-1}}\right)
 \left(\frac{T_\infty}{0.5~\mbox{keV}}\right)^{-4},
  \label{eq:Loo}
\end{equation}
which is a function of only observed quantities and is independent of
the neutron-star temperature in our model calculations. We use this
quantity as a measure of the brightness of the star at infinity and,
in Figure~5, we plot against it the maximum pulse fraction for a 10~km
neutron star emitting a blackbody spectrum. This figure shows clearly
that, even though emission from a neutron-star surface can in
principle produce pulse fractions as large as $\sim 30$\% (for very
small hot spots), it can never produce, e.g., a pulse fraction of
$\gtrsim 10$\% simultaneously with an inferred luminosity of 
$\gtrsim 10^{36}$~erg~s$^{-1}$ for a 0.5~keV blackbody temperature.

\section{DISCUSSION}

In this paper we have shown that the inferred properties of thermally
emitting neutron stars, i.e., their emitting surface areas and pulse
fractions, are significantly affected by the three-dimensional
geometry of the systems, phase-averaging effects, and general
relativistic light bending. The uncertainties introduced by these
effects can be significant (with a spread of $\lesssim 2$), especially
at the limit of small emitting surface areas. However, even when most
of the neutron star surface is emitting, the uncertainties in
estimating the neutron-star radius are of the same order as the $\sim
5$\% level required (Lattimer \& Prakash 2000) for constraining the
equation of state of neutron-star matter.

We have also argued that for a given local radiation spectrum emerging
from a bright spot on the stellar surface, faint sources can give rise
to larger pulse fractions than brighter sources. The maximum pulse
fraction as a function of the brightness of the neutron star, measured
by $l$ (cf.\ eq.~[\ref{eq:Loo}]), provides a diagnostic tool for
distinguishing between different emission models.  These derived
constraints depend both on the local radiation spectrum (e.g., thermal
versus non-thermal emission) and its beaming and are, therefore,
different for different classes of models. Moreover, both quantities
are measurable from spectral and timing observations and can be
directly compared to the calculated constraints. The high observed
pulse fractions of AXPs (see, e.g., Chakrabarty et al.\ 2000) and
X-ray bursters (Strohmayer et al.\ 1998) have been shown to strongly
constrain emission models and such constraints can become only tighter
when the apparent luminosities of these systems are also taken into
account. As an example, we consider the source RXS~J1708$-$40, which
has been identified as an AXP (Sugizaki et al.\ 1997). Fitting a
blackbody spectrum to {\em ASCA\/} observations of this source gives a
temperature of $T_\infty=0.41\pm 0.03$~keV and a blackbody flux of
$4\pi d^2 F_\infty=3.16\times 10^{35}$~erg~s$^{-1}$ (Sugizaki et al.\
1997; Chakrabarty et al.\ 2000). The inferred apparent brightness of
this source is, therefore, $l=0.8$ and combined with the observed
$\simeq 50$\% pulse fraction (Sugizaki et al.\ 1997) is inconsistent
with isotropic thermal emission from the neutron-star surface (see
Fig.~5).

Finally, our results also have a number of important implications for
soft X-ray surveys, in supernova remnants, for young, cooling neutron
stars that emit thermally.  For a given spectral shape, the brightest
sources, which would be more easily detectable, correspond to smaller
pulse fractions, while strong pulsations are only expected from dimmer
sources. This effect is shown in Figure~6, where the fraction of
systems $N_{\rm obs}/N_{\rm tot}$ with a pulse fraction at infinity
larger than a threshold $PF_{\rm 0}$ is plotted against the fractional
emitting surface area. For this purpose, we assume a random
orientation of the magnetic inclination and the inclination to the
observer for a sample of systems and define
\begin{equation}
\frac{N_{\rm obs}}{N_{\rm tot}}(PF_0) =
\int_{\alpha=0}^{\pi/2}\int_{\beta=0}^{\pi/2}
   X[PF(\alpha,\beta); PF_0] \sin\alpha d\alpha\; \sin\beta d\beta \;,
\end{equation}
where the step function $X(PF;PF_0)$ is defined such that it is unity
when $PF>PF_0$ and zero at all other values of the pulse fraction.
For realistic neutron star masses and radii ($p\sim 2-3$), only a very
small fraction of sources shows pulsations that are detectable at a
significant level ($\gtrsim 10-20$\%) and this fraction drops rapidly
with increasing apparent luminosity.

If the central source in the remnant Cas A is a young, cooling neutron
star, its surface brightness distribution cannot be uniform, as
inferred from fitting thermal models to the observed countrate spectra
(Pavlov et al.\ 2000; Chakrabarty et al.\ 2000). However, as Figure~6
shows, this property is not inconsistent with the $\simeq 30$\% upper
limit on its pulse fraction (Chakrabarty et al.\ 2000). For the
assumed isotropic beaming of the emerging radiation and the very small
fractional surface area inferred for the central source in Cas~A
(Pavlov et al.\ 2000; Chakrabarty et al.\ 2000), less than half of the
systems would have been detected with pulse fractions higher than the
detection threshold, if $p=4$. For more realistic neutron-star
properties ($p=2,3$), no system would show such a large pulse
fraction. Therefore, the absence of detectable pulsations from this
source is not a strong argument against its identification with a
spinning neutron star.

\acknowledgements

We thank Deepto Chakrabarty, Lars Hernquist, and Ramesh Narayan for
many useful discussions. D.\,P.\ acknowledges the support of a
post-doctoral fellowship from the Smithsonian Insitution. F.\,\"O.\
acknowledges support of NSF Grant AST-9820686.

\end{document}